\documentclass[aps,prl,twocolumn,groupedaddress,showpacs]{revtex4}

\usepackage[pdftex]{graphicx}

\begin{document}

\title{Determination of the Superfluid Gap in Atomic Fermi Gases by Quasiparticle Spectroscopy}

\author{Andr\'{e} Schirotzek, Yong-il Shin, Christian H. Schunck and Wolfgang Ketterle}

\affiliation{Department of Physics, MIT-Harvard Center for
Ultracold Atoms, and Research Laboratory of Electronics,
Massachusetts Institute of Technology, Cambridge, Massachusetts,
02139}

\date{\today}

\begin{abstract}
We present majority and minority radiofrequency (RF) spectra of strongly interacting imbalanced Fermi
gases of $^6$Li. We observed a smooth evolution in the nature of pairing
correlations from pairing in the superfluid region to polaron
binding in the highly polarized normal region. The imbalance
induces quasiparticles in the superfluid region even at very low
temperature. This leads to a local bimodal spectral response,
which allows us to determine the superfluid gap $\Delta$ and the
Hartree energy $U$.
\end{abstract}


\maketitle

Pairing and superfluidity in fermionic systems are intricately
related phenomena. In BCS theory \cite{bard57}, describing
conventional superconductors, the emergence of superfluidity is
accompanied by the opening of a gap in the excitation spectrum of
the superfluid. This gap can be interpreted as the minimum energy
required to break a Cooper pair or, equivalently, to create an
elementary excitation, a so-called quasiparticle, inside the
superfluid.

However, strongly correlated systems show a more complicated
behavior. There are gapless fermionic superfluid systems, as is the case for high temperature superconductors
\cite{lee06rev} or for superconductors with magnetic impurities
\cite{abri61}. On the other hand, there are numerous
examples of systems with an excitation gap in the normal state,
e.g. a high temperature superconductor above its superfluid
transition temperature exhibiting a pseudogap \cite{lee06rev} or a
semiconductor.

Here, we use radiofrequency spectroscopy to investigate the
nature of pairing and the relation between pairing and
superfluidity in a strongly interacting system of ultracold atomic
Fermions.

We can spectroscopically distinguish the superfluid and the
polarized normal fluid by introducing excess fermions into the
system. In a superfluid phase described by BCS theory, the excess particles can be
accommodated only as thermally excited quasiparticles. A double-peaked spectrum reflects the co-existence of pairs and
unpaired particles. In the normal phase, at large spin
polarization, the limit of a single minority particle immersed
into a Fermi sea is approached, which can be identified as a
polaron \cite{chev06univ,comb07,prok08pol,mass08pol}. Here the system can be described in the framework of Fermi liquid theory and no stable pairs exist. We find
that these different kinds of pairing correlations are smoothly connected
across the critical density imbalance \cite{shin08imb},
also called the Clogston-Chandrasekhar limit of superfluidity
\cite{clog62,chan62}.

Excess fermions in a low-temperature superfluid constitute quasiparticles
populating the minimum of the dispersion curve. The RF spectrum of a superfluid with such quasiparticles shows two peaks, which, in the BCS limit, would be split by $\Delta$, the superfluid gap. Therefore, RF spectroscopy of quasiparticles is a direct way to observe the superfluid gap $\Delta$ in close analogy with tunneling experiments in superconductors \cite{giaev60}. From the observed spectrum we can also determine a Hartree term \cite{cast08var}, whose inclusion turned out to be crucial.

For this study, we have combined several recently developed
experimental techniques: The realization of superfluidity with
population imbalance \cite{zwie06imb} leading to phase separation \cite{zwie06imb,part06imb,shin08imb}, tomographic RF spectroscopy
\cite{shin07rf}, in-situ phase contrast imaging with 3D
reconstruction of the density distributions \cite{shin08imb}. In order to
minimize final state effects \cite{schu08pair} we have prepared an imbalanced
mixture of states $|1\rangle$ and $|3\rangle$ of $^6$Li (corresponding to
$|F=1/2,m_{F}=1/2\rangle$ and $|F=3/2,m_{F}=-3/2\rangle$ at low
field) in an optical dipole trap at a magnetic field of $B=690$ G,
at which there is a Feshbach scattering resonance between the
states $|1\rangle$ and $|3\rangle$ \cite{schu08pair,bart04fesh}. Evaporative cooling at $B=730$ G is performed by
lowering the power of the trapping light. After equilibration an RF pulse was applied for $200$ $\mu$s selectively driving a hyperfine
transition from state $|1\rangle$ or $|3\rangle$ to state ($|2\rangle$$|F=1/2,m_{F}=-1/2\rangle$ at low
field). The RF power was kept constant for all experimental
results presented. Immediately after the RF pulse an absorption
image was taken of the atoms transferred into state
$|2\rangle$.

The spectra were correlated to the local Fermi energy
$\epsilon_{F\uparrow} =
\frac{\hbar^2}{2m}\,\left(6\pi\,n_{\uparrow}\right)^{2/3}$ of the
majority density $n_{\uparrow}$ and to the local polarization
$\sigma_{loc}=\frac{n_{\uparrow}-n_{\downarrow}}{n_{\uparrow}+n_{\downarrow}}$
which is a measure of the local excess fermion population. As in a previous publication
\cite{shin08imb} the local densities were measured using phase
contrast imaging and 3D reconstruction using the inverse Abel
transformation.

\begin{figure*}
    \begin{center}
        \includegraphics[height=1.5in]{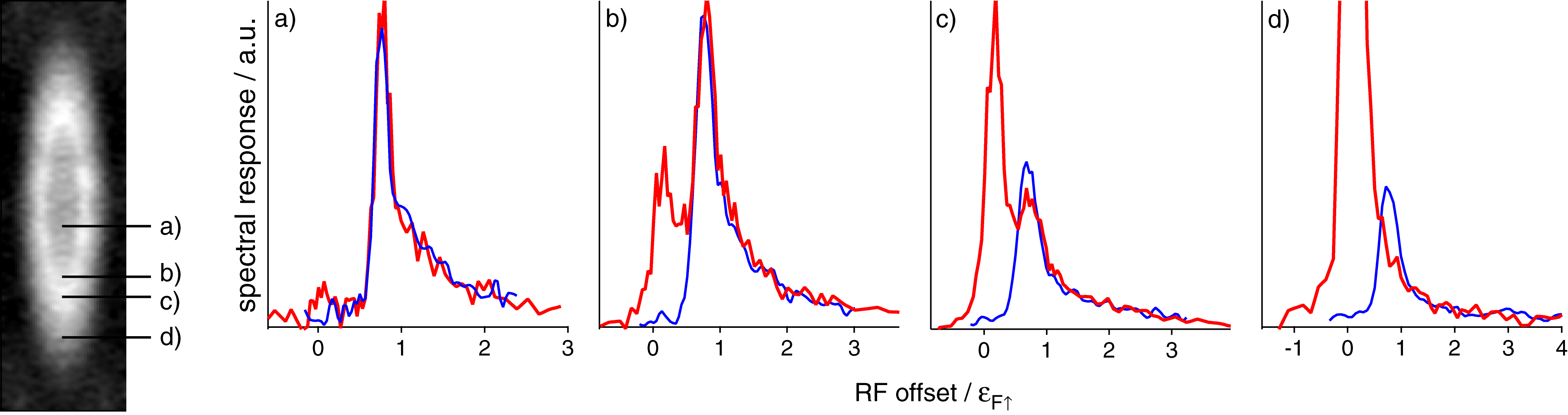}
\caption{(Color online) Tomographically
reconstructed RF spectra for various regions of the atomic sample at
unitarity. a) Balanced superfluid, b) polarized superfluid, c)
moderately polarized transition region and d) highly polarized
normal region. The panel on the left shows a phase contrast image of the atomic cloud before RF excitation. The positions of the spectra a) to d) are marked in the phase contrast image and by the arrows in figure \ref{2dspectrum} which displays all our
results in the unitary limit. Red: Majority spectrum, blue:
Minority spectrum. Local polarizations $\sigma_{loc}$ and local
temperature $T / T_F$, respectively: a) -0.04(2), 0.05(1), b)
0.03(1), 0.06(1)c) 0.19(1), 0.06(2), d) 0.64(4), 0.10(2). The negative value in a) implies that the local polarization as inferred from phase-contrast imaging underestimates $\sigma_{loc}$ by up to 0.05.
\label{4spectra}}
    \end{center}
\end{figure*}

The RF spectra shown in fig. \ref{4spectra} reveal a gradual
change in the nature of the pairing correlations. The balanced
superfluid is characterized by identical spectral responses of
majority and minority particles and has been the subject of
previous studies, see \cite{kett08var,grim07var,jin08var} and
references therein. In the polarized superfluid region
\cite{shin08imb,shee07imb} (and references therein) the minority spectrum perfectly matches the
pairing peak of the majority spectrum, locally coexisting with the
quasiparticle spectral contribution, resulting in a local double
peak structure of the majority spectrum (see fig.
\ref{4spectra}b). The spectrum suggests that the majority
population can be divided into two distinct parts: One part
consisting of pairs forming the superfluid, the other part
consisting of quasiparticle excitations in the form of excess
fermions. Therefore, a natural interpretation of the RF spectrum is to identify one peak as a Stokes process (RF creates a quasiparticle excitation) giving rise to the dissociation part of the RF spectrum and the other as an Anti-Stokes process (RF destroys a quasiparticle excitation).

\begin{figure}
    \begin{center}
        \includegraphics[height=2.5in]{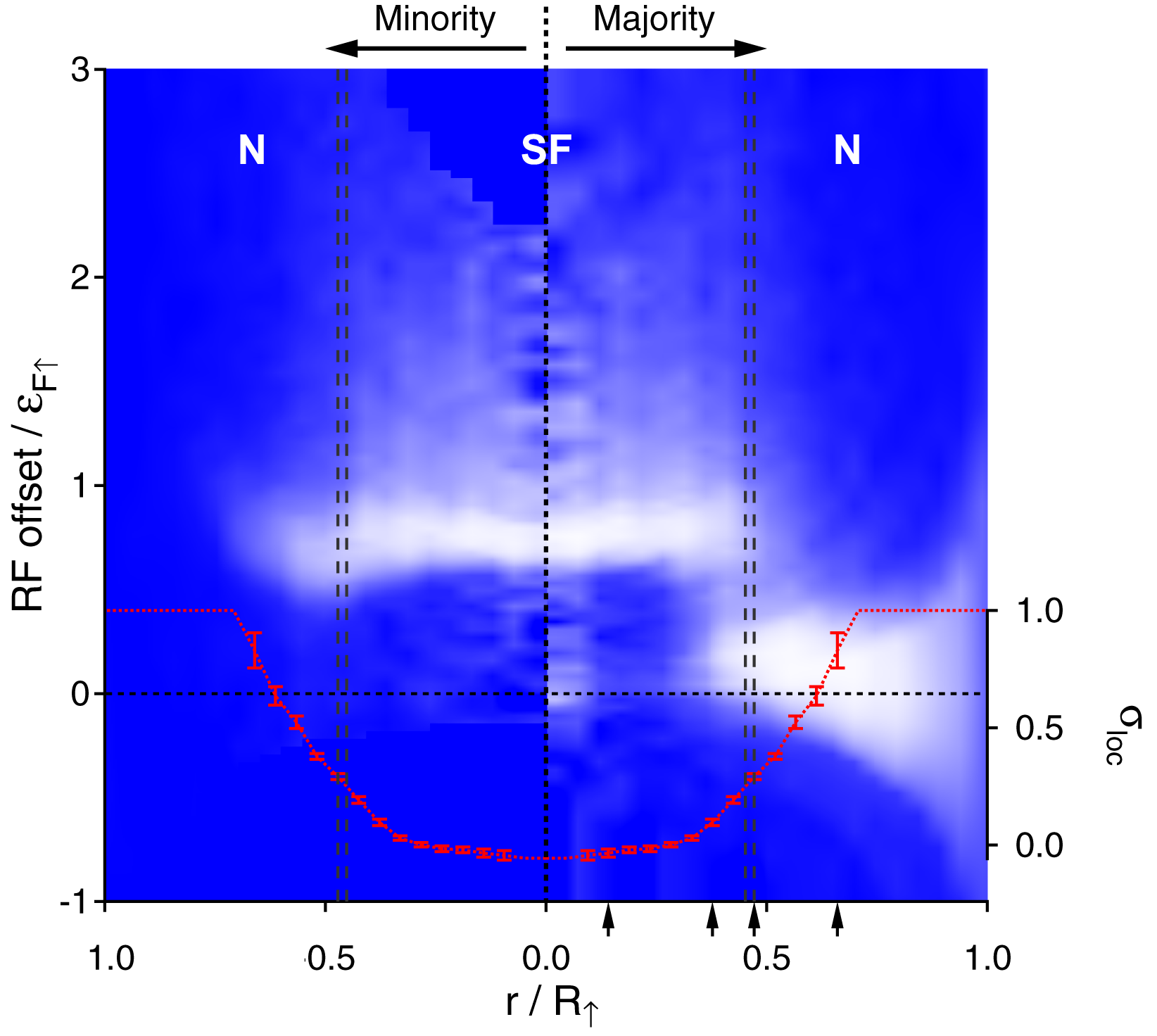}
\caption{(color online) Spatially resolved RF spectra of an
imbalanced Fermi gas at unitarity. a) The right half shows the
majority spectra as a function of position in the trap expressed
in terms of the majority Fermi radius R$_{\uparrow}$, the left
half displays the minority spectra. The superfluid to normal
transition region is marked by the gray vertical lines. The local
polarization $\sigma_{loc}$ is given by the dashed red line. The
error bars are the standard deviation of the mean value. The
arrows indicate the position of the four spectra shown in fig.
\ref{4spectra}. The image is a bilinear interpolation of 2500 data
points, each plotted data point in the image is the average of
three measured data points. The spatial resolution of the image is 0.045$\cdot
R_{\uparrow}\label{2dspectrum}$}
    \end{center}
\end{figure}

As the local imbalance is further increased beyond the superfluid
to normal (SF-N) transition \cite{epaps}, see fig. \ref{4spectra}d, the majority
spectrum no longer shows a local double-peak structure. This is
consistent with theoretical work \cite{mass08rf, muel07rf} attributing the double peak structure in the normal phase in previously reported RF spectra \cite{chin04, schu07rf} to the inhomogeneous density
distribution. For increasing spin polarization the majority and
minority pairing peaks lose spectral overlap. We interpret the missing overlap as indication that the minority atoms are no longer bound in pairs, each of them interacting with more than one majority atom, a situation we refer to as polaronic binding. We have seen \cite{epaps} that on the BEC side of the Feshbach resonance
the overlap between minority and majority spectra does not depend
strongly on the presence of excess fermions. This is expected in a
molecular picture, where the pairing character is independent of
the presence of excess fermions. At unitarity, within our
experimental resolution, the overlap starts to decrease at the
superfluid-to-normal interface, see fig. \ref{4spectra}c. This raises the question whether at unitarity a local probe
of the microscopic pairing, like RF spectroscopy, can distinguish
whether the system is superfluid or not.

Even when the spectral overlap decreases, there is still equal response
to the RF excitation in the high frequency tails, see fig.
\ref{4spectra}c and fig. \ref{4spectra}d. These tails correspond
to large momentum components in the interparticle wave function and hence address the short range physics. We expect this part of the spectrum to be insensitive to changes in the binding at large distances.

The direct comparison between majority and minority spectra
clarifies our previous experimental results on minority RF excitation spectra in the $|1\rangle - |2\rangle$ mixture
\cite{schu07rf}, in which we concluded
that there is strong pairing in the normal phase. However, the observed spectral gap in the normal phase should not be interpreted as a signature of pairing but rather as strong pairing correlations in the form of a polaron as suggested in \cite{punk07rf, veil08rf, pera08rf}. The change in pairing correlations is indiscernible in
the minority spectrum alone, but shows up in the spectral overlap
with the majority spectrum.

\begin{figure}
    \begin{center}
        \includegraphics[height=1.2in]{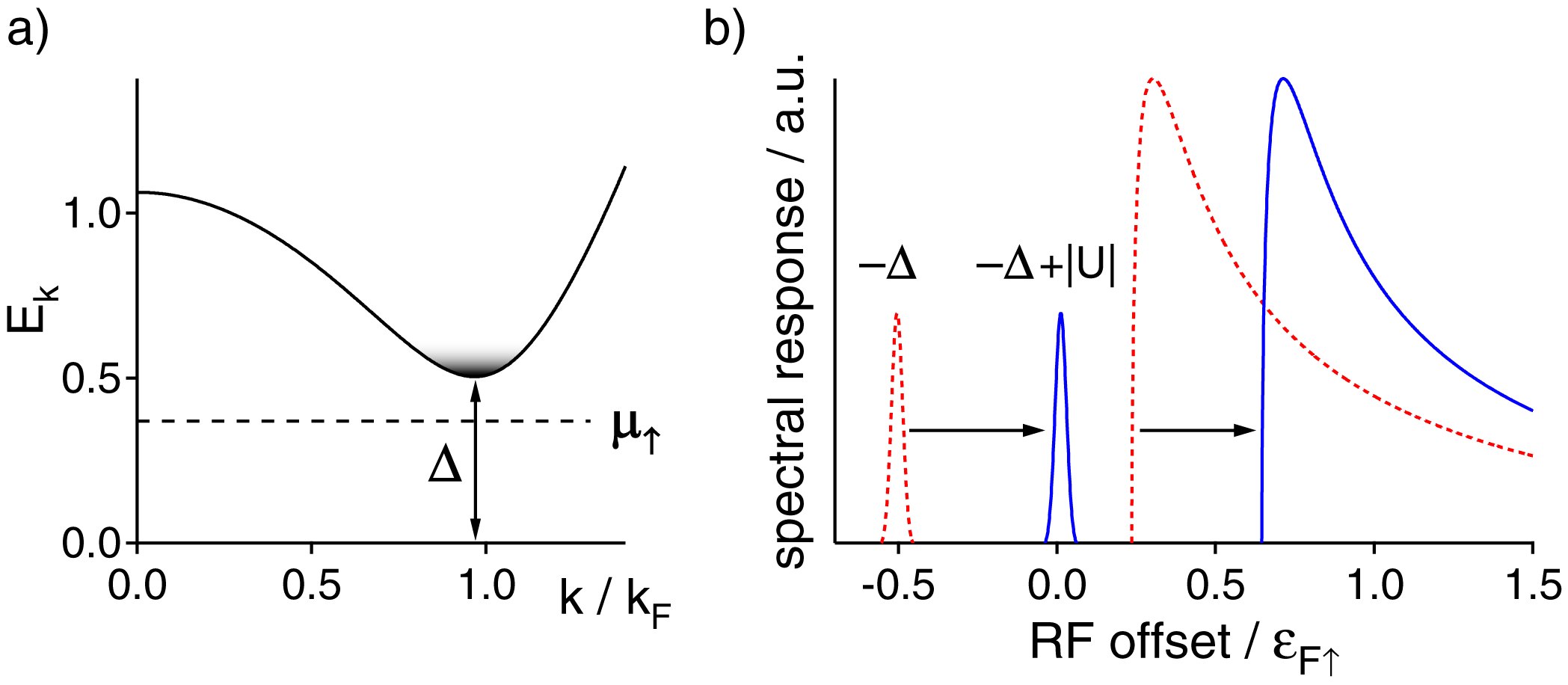}
\caption{(color online) Creation and spectroscopy of
quasiparticles . a) Population imbalance thermally generates
quasiparticles even at low temperatures comparable to
$\Delta-\mu_{\uparrow}$. $\mu_{\uparrow}$ is the chemical
potential of the majority component. b) The RF spectrum consists
of a quasiparticle peak at negative frequencies and the pair
dissociation spectrum at positive frequencies (dotted line). On
resonance, the Hartree contribution $U$ acts as an effective
attraction and hence shifts the entire spectrum into the positive
direction.\label{dispersion}}
    \end{center}
\end{figure}

We now turn to a quantitative analysis of the spectral peaks in
the superfluid phase for small density imbalance, and to the
determination of the superfluid gap. Earlier work \cite{chin04,
shin07rf} tried to determine the gap from the onset of the pair
dissociation spectrum. However, the RF spectrum is not only sensitive
to final state interactions, it is also shifted by Hartree
energies \cite{cast08var,bulg08gap}, as we show here. Furthermore, RF spectroscopy can
excite all fermions, even deep in the Fermi sea \cite{kett08var}.
Therefore, the onset of the pair dissociation spectrum occurs for
atoms with momentum $k=0$ and, in the BCS limit depends quadratically on the gap parameter ($\omega_{th}=\frac{\Delta^2}{2 \epsilon_F}$). The excitation gap can
be directly observed if quasiparticles near the dispersion
minimum are \emph{selectively} excited, as in tunnelling
experiments.

Our solution is to study not the ground state of a superfluid, but
excited states where quasiparticles are present. In a simple BCS
description, quasiparticles are in pure momentum states, but
increase the total energy of the system because their momentum
state is no longer available to the other particles for pairing.
Consequently, in an excitation spectrum, quasiparticles appear at
negative frequencies relative to the bare atomic transition
frequency. The lowest energy quasiparticle appears at frequency
$-\Delta$, see fig. \ref{dispersion}.

Final state interactions and Hartree terms can also
create line shifts, and two peaks are needed for
analysis, the dissociation peak and the quasiparticle peak in our case. In
essence, it is the separation between the peaks in spectra like
fig. \ref{4spectra}b, which allows us to determine $\Delta$.

Thermal population of quasiparticles requires a temperature on
the order of the excitation gap $\Delta$. At unitarity, this temperature can
be estimated to be $95\%$ of the critical temperature, away from the low temperature limit addressed in this letter.
Indeed, in samples of equal population of the spin
states we were not able to spectroscopically resolve any local
double peak structures \cite{epaps}. This problem can be
overcome by introducing density imbalance between the
constituents: The Fermi pressure (chemical potential $\mu_{\uparrow}$) of majority
atoms forces a finite quasiparticle occupation into the superfluid
region already at very low temperature. This allows us to
selectively populate quasiparticles at the minimum of the
dispersion curve see fig. \ref{dispersion}a.

\begin{figure}[!h]
    \begin{center}
        \includegraphics[height=1.5in]{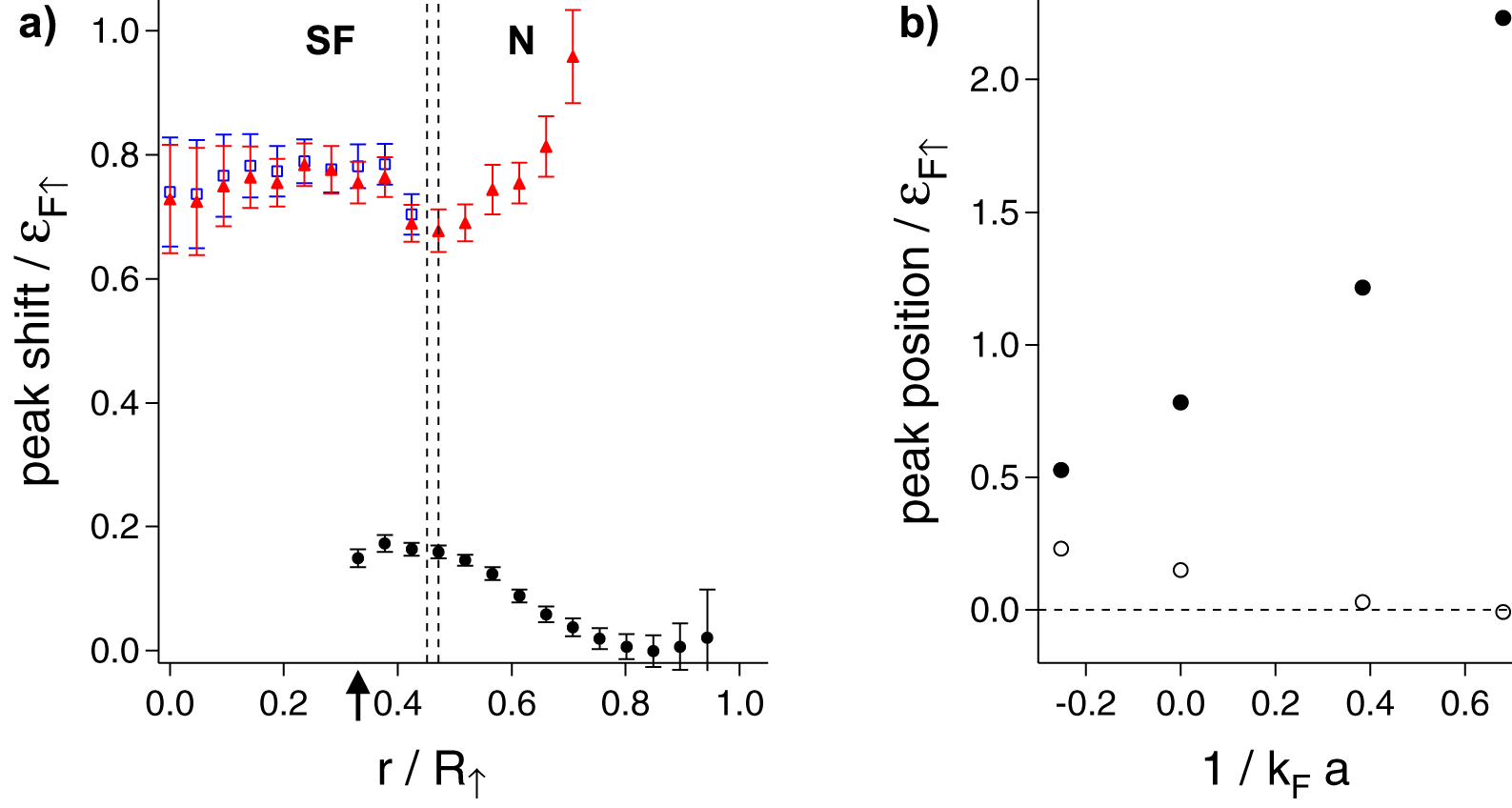}
\caption{(color online) a) Normalized peak positions of pairing
peaks and quasiparticle peak at unitarity as a function of position in the trap. The SF-N boundary is marked by the dashed vertical lines.
The arrow indicates the limit of low quasiparticle population relevant for b). Majority: blue open squares
(pairing peak) and solid black circles (quasiparticle peak),
Minority: solid red triangles. b) Pairing peak and quasiparticle peak
positions as a function of the local interaction strength
$1/k_{F} a$ in the limit of small local imbalance (see
arrow in a). Pairing peak: Solid circles,
quasiparticles peaks: Open circles.\label{peakpos}}
    \end{center}
\end{figure}

In figure \ref{peakpos}a the position of the peaks of the majority
and minority spectra are plotted normalized by
$\epsilon_{F\uparrow}$ as a function of position in the trap for
the unitary limit \cite{epaps}. The peak positions are proportional to the
local Fermi energy inside the superfluid region within our
experimental resolution. In the region of superfluidity with
finite polarization the spectra show local double peaks. The position of the two peaks in the limit of small polarization is depicted in fig. \ref{peakpos}b for various interaction strengths.

It was unexpected that the quasiparticles appear at
positive frequencies (relative to the atomic transition
frequency). This is caused by the presence of Hartree terms
\cite{cast08var}, resulting in an overall shift of the systems energy and
the RF spectrum \cite{epaps}. In the weakly interacting limits, the Hartree term
reduces to a simple mean field shift. In
the strongly interacting regime one has to resort to QMC
calculations \cite{carl05imb, lobo06imb, bulg08gap} for a
numerical value of U.

In a mean-field description of the balanced superfluid starting from
the BCS-Leggett ansatz for the BEC-BCS crossover \cite{kett08var} taking into
account the Hartree term $U$, the dispersion relation of the
quasiparticles can be expressed as $E_k=\sqrt{\Delta^2+\left(\epsilon_k+U-\mu\right)^2}$ \cite{cast08var}
where $\epsilon_k = \frac{\hbar^2 k^2} {2 m}$ is free particle kinetic
energy and $\mu$ is the chemical potential. This mean-field formalism gives
the analytic expression for the two peak positions. A quasiparticle at the
minimum of the dispersion curve will respond at an RF offset of $\omega_{RF}=-E_{k_{min}}-\mu+\epsilon_{k_{min}}=-\Delta-U$, and
the maximum of the pair dissociation spectrum occurs at $\hbar \omega_{max}=\frac{4}{3} \left(\sqrt{\mu'^2+\frac{15}{16} \Delta^2}-\mu'\right)-U \simeq \frac{4}{3}\,
\omega_{th}-U$, where $\mu'=\mu - U$ and $\omega_{th}$ is the
dissociation threshold (which is at momentum $k=0$).

We determined the superfluid gap $\Delta$ and the Hartree energy $U$ from the peak
positions in the limit of small density imbalance ($\sigma_{loc} \simeq 0.03$). At
unitarity with the chemical potential $\mu=0.42 \,\epsilon_{F\uparrow}$,
confirmed in previous experiments, see \cite{haus07} and references therein, we obtained $\Delta=0.44
\,\epsilon_{F\uparrow}$  and $U=-0.43 \,\epsilon_{F\uparrow}$, in excellent agreement with the predicted values
$\Delta_t=0.45 \,\epsilon_{F\uparrow}$ and $U_{t}=-0.43 \,\epsilon_{F\uparrow}$ from QMC calculations \cite{carl08gap}. Our
determined values for $\Delta$ and $U$ values suggest the minimum of the quasiparticle
dispersion curve to occur at $k_{min} \simeq 0.9 k_F$. Table \ref{table} shows the gap and
Hartree energy for various interaction strengths. Away from unitarity we
relied on QMC calculations for the chemical potential $\mu$ \cite{astr04eos}.

For an accurate quantitative comparison \cite{epaps} final state interactions,
also listed in table \ref{table}, had to be taken into account.
The effect of final state interactions is an overall mean
field shift of $E_{final}=\frac{4 \pi \hbar^2 a}{m} n$. This shift affects both the quasiparticle peak and
the pairing peak equally.

\begin{center}
    \begin{table}
    \caption{Superfluid gap $\Delta$, Hartree term $U$ and final state interaction $E_{final}$ in terms of the Fermi energy $\epsilon_{F\uparrow}$ for various interaction strengths $1 / k_F a$.\label{table}}
        \begin{tabular*}{0.3\textwidth}%
      {@{\extracolsep{\fill}}ccccc}
        \hline\hline
        $1 / k_F a$ & $\Delta$ & $U$ & $E_{final}$\\\hline
        -0.25 & 0.22 & -0.22 & 0.22\\
        \textbf{0} & \textbf{0.44} & \textbf{-0.43} & \textbf{0.16}\\
        0.38 & 0.7 & -0.59 & 0.14\\
        0.68 & 0.99 & -0.87 & 0.12\\\hline\hline
        \end{tabular*}
    \end{table}
\end{center}

In conclusion, we have performed spatially resolved RF spectroscopy of majority and minority components of a trapped
imbalanced Fermi gas in the strongly interacting regime with small
final state effects. In crossing the superfluid to normal boundary
we observed a gradual crossover in the pairing mechanism by
comparing majority and minority spectra. The majority spectrum
shows a local double peak spectrum in the polarized superfluid
region which allowed us to determine the superfluid gap $\Delta$
and the Hartree terms $U$. The spectra in the normal phase are
consistent with a polaron picture.

We thank W. Zwerger and M. Zwierlein for stimulating discussions and A. Sommer for critical reading of the manuscript. This work was supported
by the NSF and ONR, through a MURI program, and
under ARO Award W911NF-07-1-0493 with funds from
the DARPA OLE program.

\section*{Auxiliary Material: Determination of the Superfluid Gap in Atomic Fermi Gases by Quasiparticle Spectroscopy}

\section*{Determination of the superfluid boundary}

\begin{figure*}
    \begin{center}
        \includegraphics[width=7in]{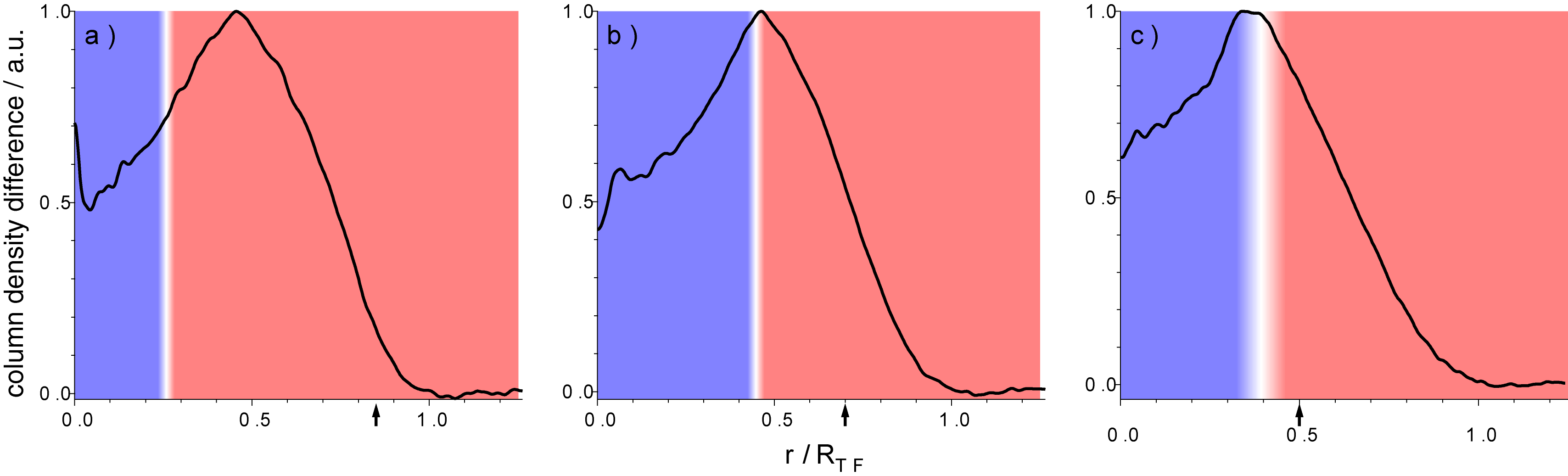}
\caption{(Color online) The difference column density profile
$n_{\uparrow}(r) - n_{\downarrow}(r)$ (radially averaged). a) BCS
side ($B=710G$), b) Unitary limit ($B=690G$), c) BEC side
($B=671G$). Each profile shown is the average of 10 individual
profiles. Blue marks the region of complete spectral overlap
between majority and minority components, red marks the region
where there is no complete spectral overlap. The black arrows at
the bottom indicate the radial size of the minority
component.\label{colDiff}}
    \end{center}
\end{figure*}

It has been shown previously \cite{shin08imb} that at unitarity
the difference column density profiles serve as an indicator for
the SF-N boundary. The discontinuity of the minority density
results in a pronounced "cusp" in the difference profile
$n_{\uparrow}(r) - n_{\downarrow}(r)$, see fig. \ref{colDiff}. In
the main body of the paper the phase boundary has been determined
by this peak position. Hence, "normal" refers to spatial regions
beyond the peak, "superfluid" refers to spatial regions inside.

The color coding in the graphs in fig. \ref{colDiff} shows where
the spectral overlap (definition see below) between the majority and minority pairing
peaks is lost. On resonance ($B=690$ G) this position shows
excellent agreement with the position of the cusp in the
difference density profile and can therefore serve as an
alternative indicator for the SF-N boundary. This coincidence
breaks down away from resonance: On the BCS side ($B=710$ G) the
spectra are less "robust" against polarization and spectral
overlap is lost before the column density difference shows a peak.
The reverse situation occurs on the BEC side of the resonance
($B=671$ G). Note that on the BEC side the minority cloud does not
extend much further than the peak position in the column density
difference.

\section*{Quantification of spectral overlap}

\begin{figure}
    \begin{center}
        \includegraphics[width=2.0in]{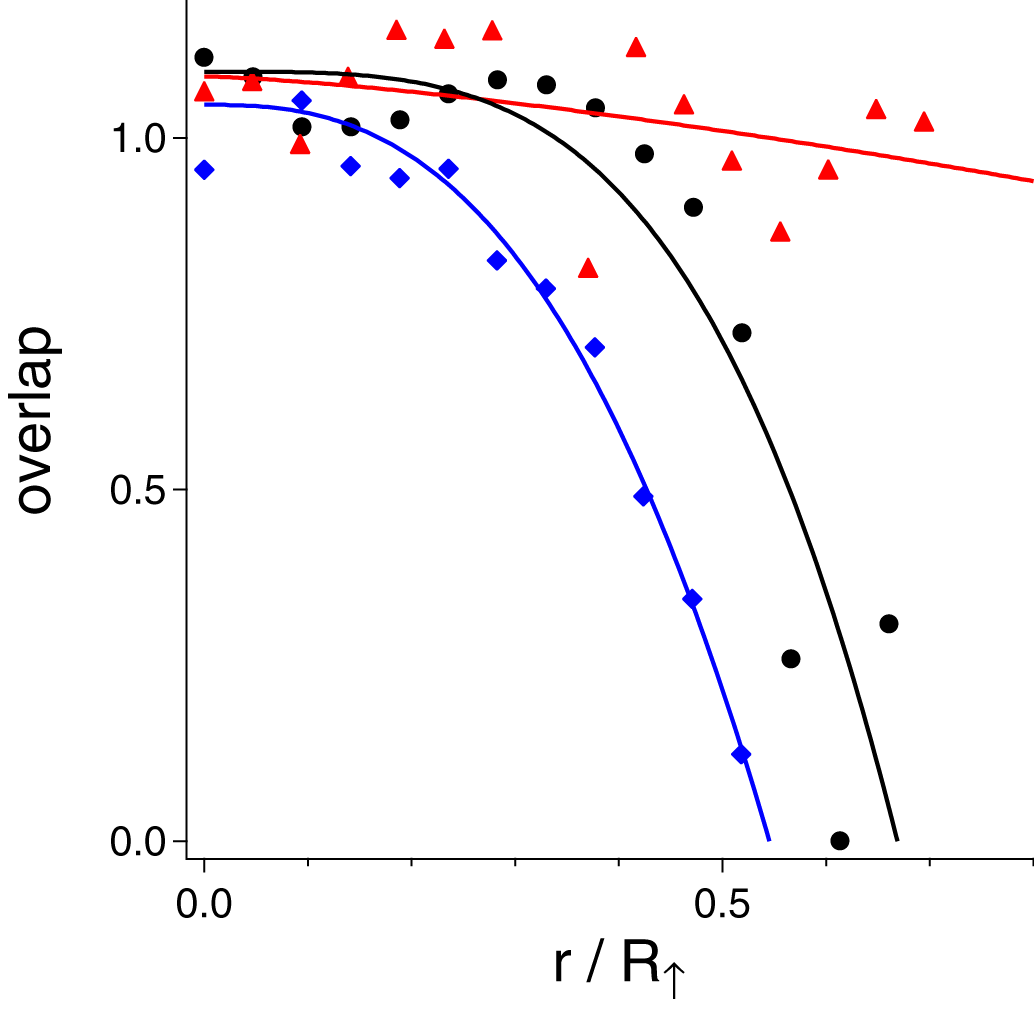}
\caption{(color online) Overlap of majority and minority
pairing peak as a function of position in the trap for various
interaction strengths.  A power law was fitted to the curves as a guide to the eye. Unitary limit
(black circles): $1/k_{F\uparrow} a = 0$, phase boundary at $r_c /
R_{\uparrow}\simeq0.46$; BEC side (red triangles):
$1/k_{F\uparrow} a = 0.39(1)$, $r_c / R_{\uparrow}\simeq0.45$; and
BCS side (blue diamonds): $1/k_{F\uparrow} a = -0.25(1)$, $r_c /
R_{\uparrow}\simeq0.35$\label{ovANDas}}
    \end{center}
\end{figure}

In order to calculate the spectral overlap of a spectrum, the
quasiparticle peak was fitted by a gaussian and subtracted
from the majority spectrum. The overlap is then defined as one minus the difference of the integrated spectra normalized by the sum of the integrated spectra, see fig. \ref{ovANDas}. As mentioned in the main text and above, on the BEC side of the Feshbach resonance almost complete spectral overlap can be observed into the normal region. On the BCS side the reverse situation occurs.

\section*{Minority peak shift for high imbalance}

\begin{figure}[!h]
    \begin{center}
        \includegraphics[width=2.5in]{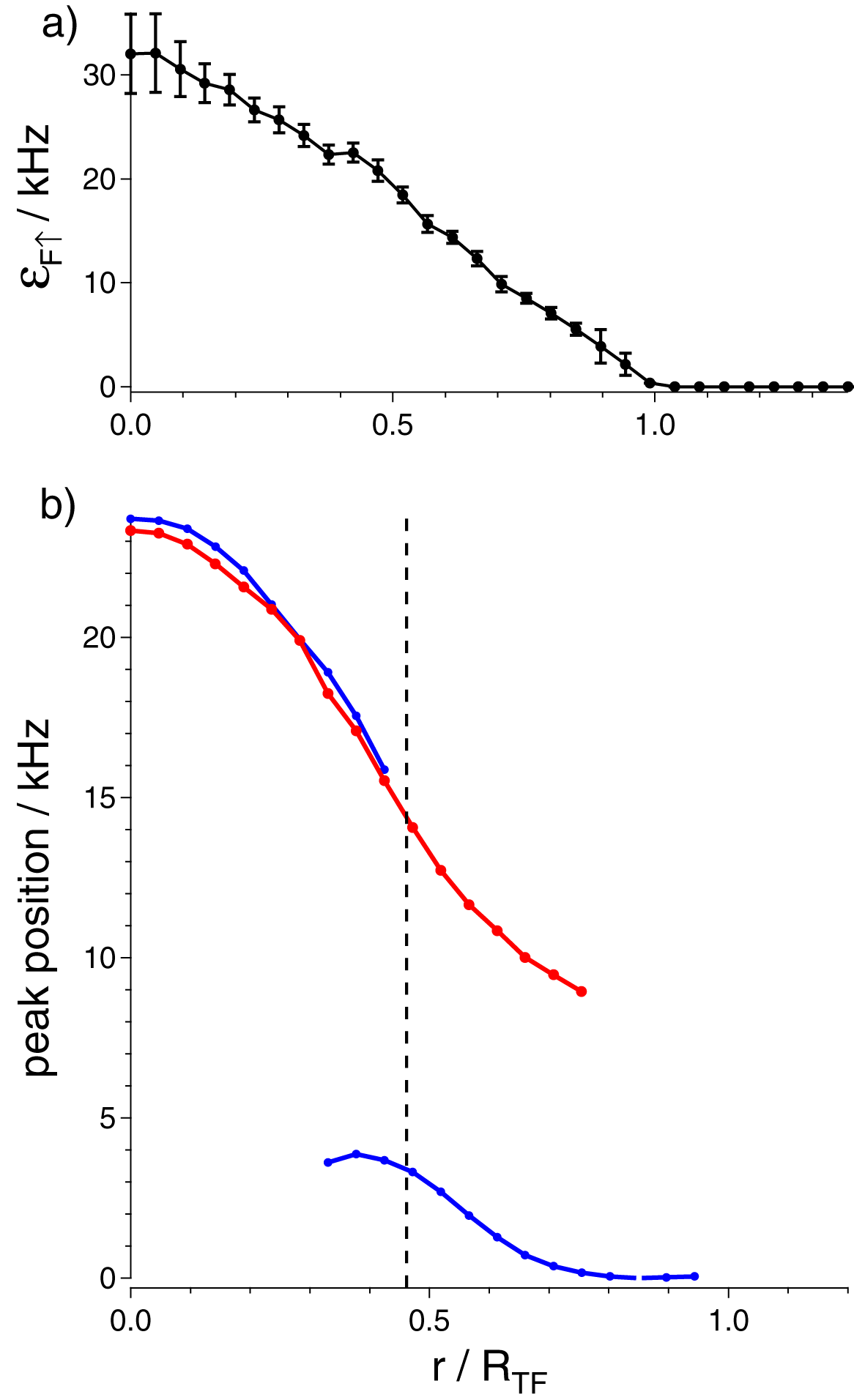}
\caption{(color online) a) Local majority Fermi energy in kHZ, the
error bars are the standard deviation of the mean value. b) Peak
positions of majority and minority in kHz. Majority (Blue):
Pairing peak (higher frequencies, only discernible in the SF
region) and quasiparticle peak (lower frequencies). Minority
(red): The peak can be traced well into the normal
region.\label{bare}}
    \end{center}
\end{figure}

Fig. 4 in the main body of the text shows the peak positions
normalized by the local majority Fermi energy $\epsilon_{F\uparrow}$.
Fig. \ref{bare} shows the bare peak positions as observed in the
experiment.

One unexpected finding in fig. 4a in the main body of the text is
the sudden increase of the minority peak position for high
imbalances. This behavior can be traced back to the data in fig.
\ref{bare}: The minority peak position (red) shows a change in
curvature as the SF-N boundary is crossed, in contrast to the majority Fermi
energy. Therefore, the ratio of minority peak and majority Fermi
energy shows a sudden increase towards the edges of the cloud. The
value of the peak position of $\omega_{pol}\simeq0.9\, \epsilon_{F\uparrow}$ (where
final state interactions of $E_{final}\simeq 0.05\, \epsilon_{F\uparrow}$ have been
taken into account) is higher than the theoretically calculated
value of $\simeq0.6\, \epsilon_{F\uparrow}$ \cite{chev06univ}.

\section*{Quasiparticles in an equal mixture}

\begin{figure*}
    \begin{center}
        \includegraphics[width=5in]{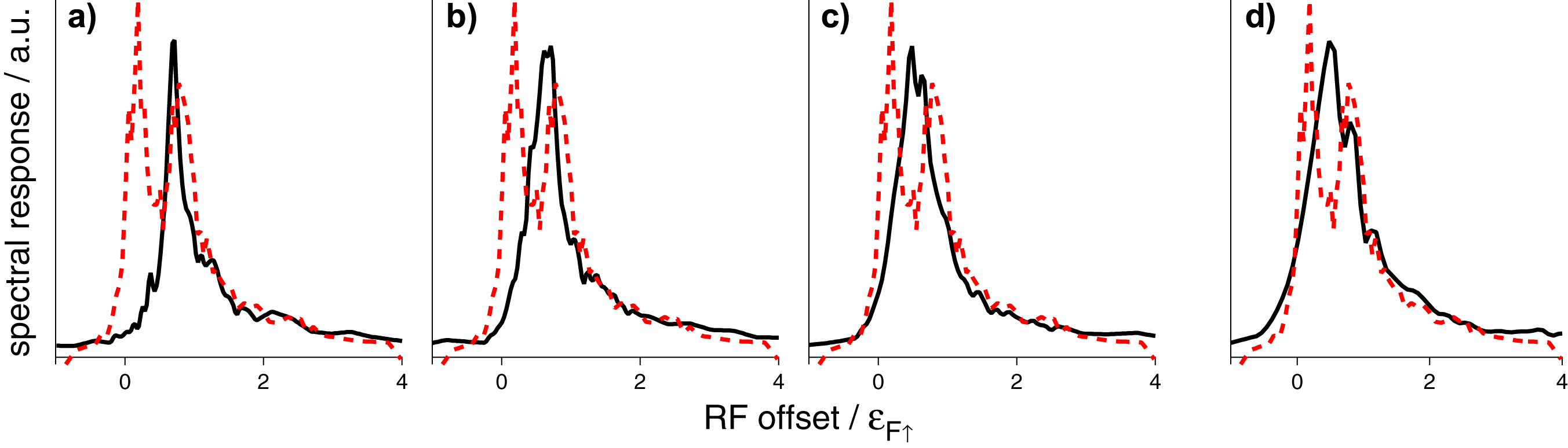}
\caption{(Color online) Local RF spectra of an equal spin mixture
for various normalized local temperatures $T / T_{F\uparrow}$. a) $T /
T_{F\uparrow}\simeq 0.20$, b) $T / T_{F\uparrow}\simeq 0.22$, c) $T /
T_{F\uparrow}\simeq 0.34$, d) $T / T_{F\uparrow}\simeq 0.55$. No local
double peak spectrum can be resolved in the RF spectrum. For
comparison, the double peak spectrum of an imbalanced mixture with
$T / T_{F\uparrow}\simeq 0.06$ is added with a red dashed
line.\label{equalQP}}
    \end{center}
\end{figure*}

In a previous publication \cite{schu08pair} low temperature RF
spectra did not show any signatures of quasiparticles. We have
attempted to create thermally generated quasiparticles in an equal
density mixture for higher temperatures $T / T_F \geq 0.20$. The
experimental results in fig. \ref{equalQP} show a decrease of the
gap parameter, but no local double peak structures could be
resolved for any temperature.

We can estimate the temperature required to populate
quasiparticles, assuming that the temperature has to be on the
order of the gap and that $\Delta(T)$ is given by the BCS relation
$\frac{\Delta(T)}{\Delta_0}=1.74 \sqrt{1-\frac{T}{T_c}}$ \cite{tink75}. At
unitarity, $\Delta_0 / T_c\simeq 3$. One would expect to populate
quasiparticles only very close to the transition temperature $T /
T_c \simeq 0.95$, when the gap is only one third of its
low-temperature value.

\section*{The theoretical dissociation spectrum including the Hartree term $U$}
Starting from a BCS-Leggett mean-field wavefunction and applying
Fermi's Golden Rule, the RF spectrum can be described by
(neglecting the Hartree term):

\begin{equation}
    \label{eq:spectrum}
    \Gamma\left(\omega\right)\propto\frac{\sqrt{\omega-\omega_{th}}}{\omega^2}\sqrt{1+\frac{\omega_{th}}{\omega}+\frac{2\,\mu}{\omega}}
\end{equation}

where $\omega_{th}=\sqrt{\Delta^2+\mu^2}-\mu$ is the dissociation
threshold and $\mu$ is the chemical potential.

The corresponding quasiparticle dispersion relation is:
\begin{equation}
    E_k=\sqrt{\Delta^2+(\epsilon_k-\mu)^2}
\end{equation}
$\epsilon_k$ being the free particle dispersion
$\epsilon_k=\frac{\hbar^2 k^2}{2m}$.

Hartree terms $U$ modify the quasiparticle excitation spectrum
\cite{cast08var,bulg08gap}:
\begin{equation}
    E_k=\sqrt{\Delta^2+(\epsilon_k-(\mu-U))^2}
\end{equation}
resulting in an RF spectrum of \begin{equation}
    \Gamma\left(\omega'\right)\propto\frac{\sqrt{\omega'-\omega_{th}'}}{\omega'^2}\sqrt{1+\frac{\omega_{th}'}{\omega'}+\frac{2\,\mu'}{\omega'}}
\end{equation}
where $\omega'=\omega+U$, $\omega_{th}'=\omega_{th}+U$ and
$\mu'=\mu-U$. This demonstrates that the spectrum retains its
functional form but the entire spectrum is shifted by $U$.

\section*{Resolution / Experimental broadening}

\begin{figure}
    \begin{center}
        \includegraphics[width=2.5in]{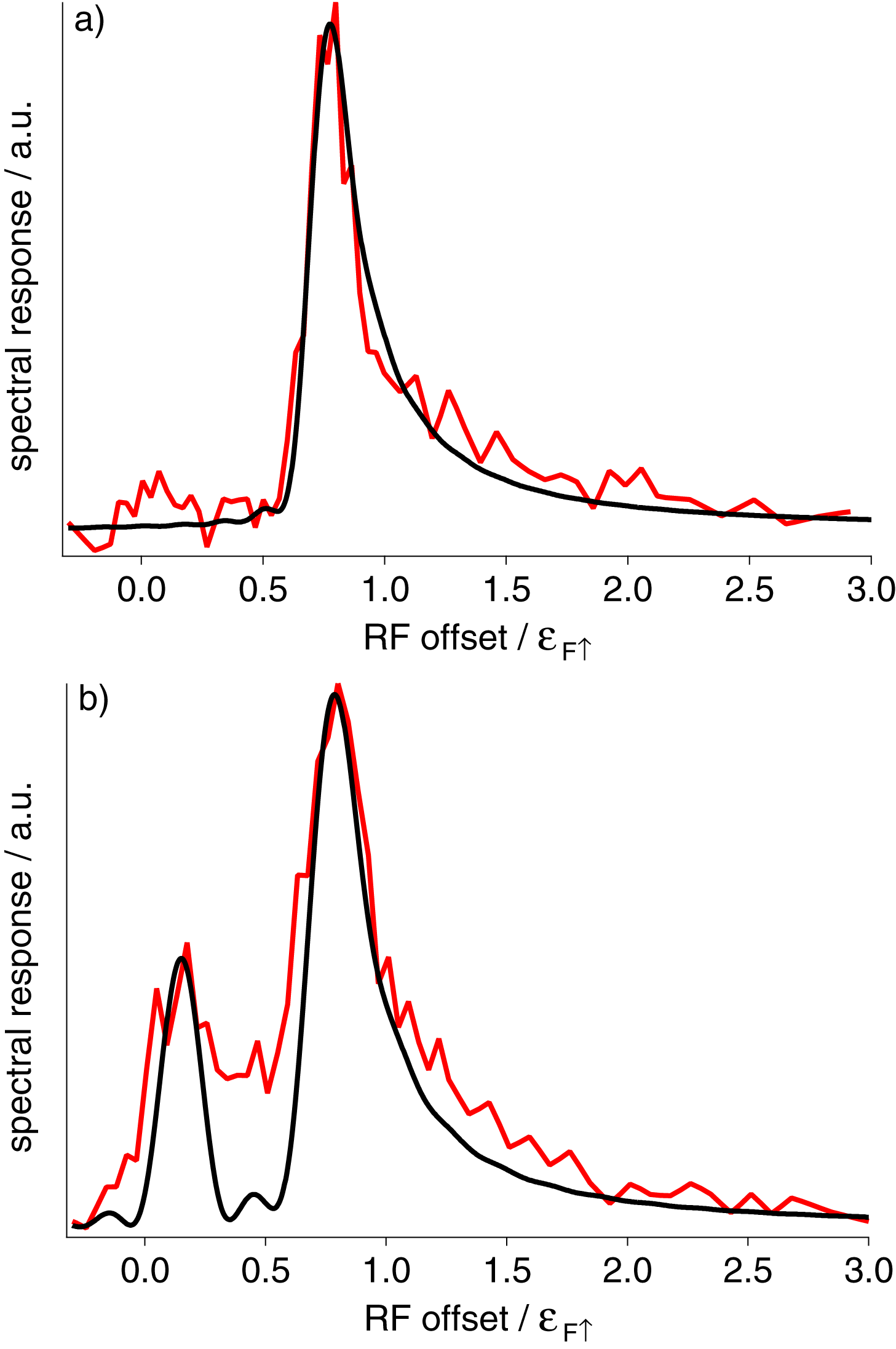}
\caption{(color online) Comparison of experimental (red) and
theoretical (black) line shapes for spectra a) and b) of figure 1
of the main body of the paper.  The theoretical curve is obtained
from a BCS-Leggett mean field description including the Hartree
term and a convolution with the experimental resolution of $\simeq
4.4$ kHz.  The values for $\Delta$ and $U$ as calculated from the
peak positions lead to a reasonable agreement with the
data.\label{broadening}}
    \end{center}
\end{figure}

For comparison with the theoretical spectrum, we model the RF pulse of $T = 200$ $\mu$s length as a square pulse,
in the frequency domain, resulting in a FWHM of the RF spectral
power of $\Delta\nu=2\cdot\frac{1}{2\pi}\,1.39 \,
\frac{2}{T}\simeq 4.4$ kHz. The theoretical spectrum consists of two parts: 1) A dissociation peak including the Hartree energy, described by eqn. \ref{eq:spectrum} with $\Delta$ and $U$ as given in table I in the main text. 2) A quasiparticle peak modeled as a narrow (FWHM $=1$ kHz) Lorentzian with a peak height adjusted so that it resembles our data. This spectrum was convolved with the Fourier transform of a square
pulse $f(\omega)\propto \frac{\sin^2{\frac{\omega
T}{2}}}{\left(\frac{\omega T}{2}\right)^2}$. Fig. \ref{broadening} shows that the
theoretical spectrum reproduces our data quite well. The deviation in
\ref{broadening}b) might be attributed to additional broadening mechanisms like finite quasiparticle lifetime, finite temperature and
atomic diffusion during the duration
of the RF pulse. The convolution causes a small shift of $0.05\, \epsilon_{F\uparrow}$ in the spectral peak position due to the asymmetry of the
theoretical spectrum and has been accounted for in the determination
of $\Delta$ and U.

The value given above for the experimental resolution is confirmed by looking
at the blurring of the sharp onset of pair dissociation. Equation
\ref{eq:spectrum} predicts that the threshold and peak position in
the strongly interacting regime differ by less than $10\%$ of the
Fermi energy. Adjusting the experimental resolution to $\sim
4$ kHz accounts for the experimentally observed difference of
threshold and peak position of about $0.3\, \epsilon_{F\uparrow}$.

\bibliographystyle{apsrev}
\bibliography{schirotzekRef}

\end{document}